\documentclass[10pt,aps,pra,twocolumn,superscriptaddress,floatfix,showpacs,longbibliography]{revtex4-1}
\usepackage[utf8]{inputenc}
\usepackage[T1]{fontenc}
\usepackage{graphicx}
\usepackage{tabularx}
\usepackage[usenames,dvipsnames,table]{xcolor}
\usepackage[version=3]{mhchem}
\usepackage{braket}
\usepackage{upgreek}
\usepackage{hyperref}
\usepackage{amsmath, amssymb}
\usepackage[normalem]{ulem}
\usepackage{soul}
\usepackage{bbold}

\definecolor{mark}{rgb}{0.85, 0.9, 1}

\hypersetup{hidelinks}
\sethlcolor{mark}

\begin{document}

\title{Quantum-optical reset with classical memory}

\author{Evgeniy O. Kiktenko}
\affiliation{Russian Quantum Center, Skolkovo, Moscow 121205, Russia}
\author{Oleg M. Sotnikov}
\affiliation{Theoretical Physics and Applied Mathematics Department, Ural Federal University, Ekaterinburg 620002, Russia}
\affiliation{Russian Quantum Center, Skolkovo, Moscow 121205, Russia}
\author{Ilia A. Iakovlev}
\affiliation{Theoretical Physics and Applied Mathematics Department, Ural Federal University, Ekaterinburg 620002, Russia}
\affiliation{Russian Quantum Center, Skolkovo, Moscow 121205, Russia}
\author{Yuri A. Biriukov}
\affiliation{Quantum Technology Centre and Faculty of Physics, M. V. Lomonosov Moscow State University, Moscow 119991, Russia}
\author{Aleksey K. Fedorov}
\affiliation{Russian Quantum Center, Skolkovo, Moscow 121205, Russia}
\author{Stanislav S. Straupe}
\affiliation{Russian Quantum Center, Skolkovo, Moscow 121205, Russia}
\affiliation{Quantum Technology Centre and Faculty of Physics, M. V. Lomonosov Moscow State University, Moscow 119991, Russia}
\author{Ivan V. Dyakonov}
\affiliation{Quantum Technology Centre and Faculty of Physics, M. V. Lomonosov Moscow State University, Moscow 119991, Russia}
\author{Vladimir V. Mazurenko}
\affiliation{Theoretical Physics and Applied Mathematics Department, Ural Federal University, Ekaterinburg 620002, Russia}
\affiliation{Russian Quantum Center, Skolkovo, Moscow 121205, Russia}

\date{\today}

\begin{abstract}
Dynamic quantum circuits generate states that depend on the measurement results obtained during circuit execution. To date such a quantum computing model has mainly been implemented with qubit-based superconducting hardware utilizing $\sf reset$ operations and classical logic. Here we develop a model of optical $\sf reset$ by using time-bin self-looped interferometers demonstrated in recent experiments. Synchronizing the optical $\sf reset$ with a simple classical device storing history of measurement results allows one to decrease uncertainty of future measurements, which suggests new possibilities for constructing dynamical circuits on optical platforms. Information flow with significant multi-time correlations and memory depth is identified through distinct information-theoretic measures. We discuss potential applications of the proposed reset model, including the realization of boson sampling and experimental tests of the quantum-mechanical formulation of Landauer’s principle. 
\end{abstract}

\maketitle

\section{Introduction}
$\sf Reset$ is one of the basic operations in quantum computing~\cite{DiVincenzo, Quantum_Supercomputer}, which enables the preparation of a quantum system in a standard state, for instance by mapping all the qubits into the $\ket{0}$ state in the $\sigma^z$ basis. Using $\sf reset$ is not limited to initialization of a quantum computing system, it is also considered an integral part of various algorithms such as iterative phase estimation~\cite{IPEA1, IPEA2, IPEA3, IPEA4}, quantum autoencoders~\cite{QAE1,QAE2}, preparing entangled states using engineered dissipation~\cite{dissipation1}, quantum error correction~\cite{QEC1, QEC2}, exploring measurement-induced entanglement~\cite{entanglement-induced} and others. When implementing the aforementioned quantum algorithms one typically employs conditional $\sf reset$, which assumes that a part of the quantum system is first measured and then, depending on either the collapsed state coincides with target one or not, additional actions (for instance, bit-flip operations) are applied, which defines a distinct trajectory in the evolution of the whole quantum system. It is worth mentioning that alternative types of $\sf reset$ have been likewise reported~\cite{reset1, reset2, reset3}.

Technically, performing time-delayed quantum operations that depend on the results of intermediate measurement has been mainly demonstrated in the case of superconducting quantum computing platforms~\cite{superconductor_reset1, superconductor_reset2, IPEA4}. These results have revealed a crucial role of implementing efficient ways for processing classical information as well as accurate synchronization between classical and quantum hardware~\cite{IPEA4}. Quantum states prepared with such $\sf reset$-based circuits are conditioned on measurement outcomes, which can provide a flexibility when one solving challenging tasks in quantum computing. For instance, it facilitates preparing projected ensembles of qubit wavefunctions~\cite{proj_ensemble1, proj_ensemble2} that feature random statistical properties. At the same time, characterizing properties of an ensemble of $2^n$ conditional qubit wave functions that can be generated using dynamical circuits with $n$ $\sf reset$s requires developing effective classical protocols for decoding measurement results~\cite{entanglement-induced}. It is natural to expect a higher complexity of monitoring conditional states in the case of multi-level (qudits) quantum systems~\cite{qudit, qudit_RQC}.    

In this work, we show that the $\sf reset$ operation, which enables the generation of conditional multi-level bosonic states, can be implemented with interferometers having a loop-based time-bin architecture~\cite{Exp1, Exp2, Exp3,Exp4,Exp5, Exp6}. For demonstration we use a minimal setup that is visualized in Fig.~\ref{loop_BS} and has dual-mode input and dual-mode output. At each step of the system's evolution, one-photon state, $\ket{1}$ is constantly prepared in the input mode $a$, entangled with a quantum state of the loop subsystem (modes $b$ and $d$) through a unitary transformation $U$, and the obtained quantum state is partially collapsed being measured with the detector in the output mode $c$. In this setting the loop provides a channel for transmitting information between ${\sf reset}$ operations performed at different times. Being synchronized with a classical memory the optical $\sf reset$ allows to decrease uncertainty of the $\sf reset$ outcome at a given time step by recording history of measurement results at the previous cycles. This clearly distinguishes our $\sf reset$ model from previous superconducting qubit realizations and paves the way to control generated conditional wave functions in real time. We show that uncertainty reduction is intimately related to information flow in the quantum device and develop distinct non-local-in-time measures to quantify it at both classical and quantum levels.


\section{Time-bin loop-based interferometer}

\begin{figure*}
  \centering
  \includegraphics[width=0.93\linewidth]{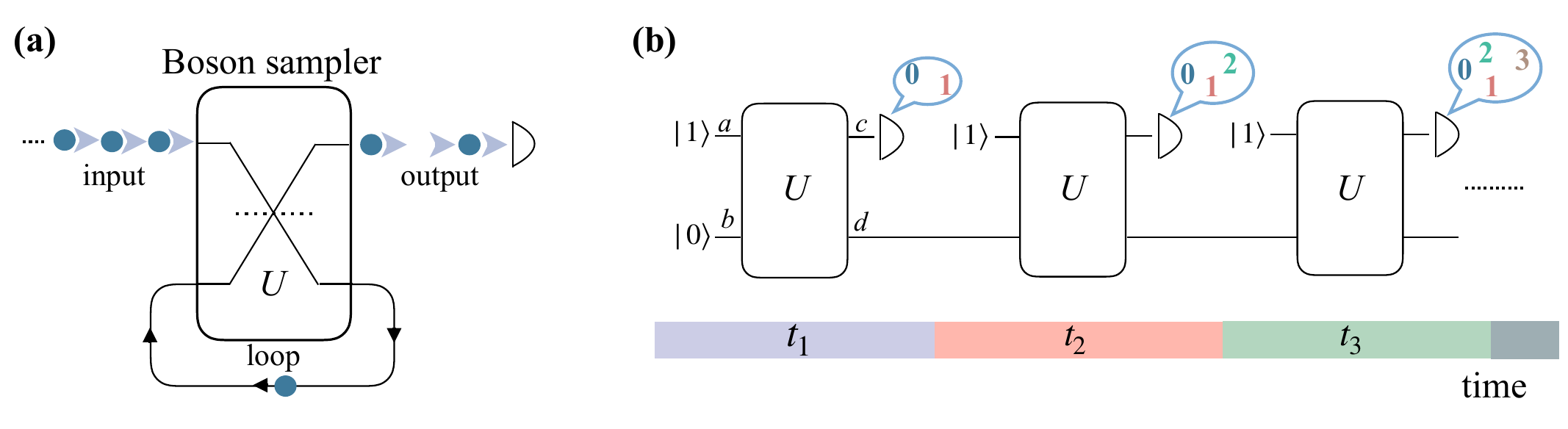}
  \caption{(a) Schematic representation of the simplest loop-based time-bin interferometer implementing the ${\sf reset}$ function. From one input mode a single photon is loaded at regular time intervals. The input and output modes are connected with a scattering unitary matrix $U$. When measured on one mode at each iteration, the output quantum state of the bosonic sampler partially collapses. The loop allows transmitting this quantum state from the previous time step to load it simultaneously with an incoming single photon at the next iteration. (b) Time-resolved decomposition of the loop-based interferometer presented in the panel (a). Each time step $t=1,2,3\ldots$ is characterized by a distinct set of events that can be detected using a photon-number-resolving detector and stored in a classical memory.
  }
  \label{loop_BS}
\end{figure*}
We start with the analysis of the quantum state time evolution of the loop-based interferometer presented in Fig.~\ref{loop_BS}. At the first time step $t=1$ the input quantum state of the interferometer is simply the tensor product of two single-mode Fock states, vacuum wave function (the empty input mode $b_1$) and single-photon state (the input mode $a_1$), $\ket{\psi^{\rm in}} = \ket{1}_{a_1} \ket{0}_{b_1}$, as shown in Fig.~\ref{loop_BS}(b). 
Hereinafter, we use subscripts within the notation of a state to specify modes to which the state belongs. 
$a_t$ ($b_t$, $c_t$, $d_t$) with $t=1,2,\ldots$ denotes mode $a$ ($b$, $c$, $d$) at the $t$-th time step as well as a corresponding annihilation operator.
The transformation of the quantum state between input and output modes is described by a $2 \times 2 $ unitary scattering matrix $U$, which relates creation operators of particles in the input ($a_t^\dagger$, $b_t^\dagger$) and output ($c_t^\dagger$, $d_t^\dagger$) modes as
\begin{equation} \label{eq:interferometer}
    \begin{aligned}
        a_t^\dagger \mapsto u_{11}c_t^\dagger + u_{12} d_t^\dagger,  \\
        b_t^\dagger \mapsto u_{21}c_t^\dagger + u_{22} d_t^\dagger,
    \end{aligned}
\end{equation}
where $u_{11}, u_{12}, u_{21}, u_{22}$ are the elements of the scattering matrix. If all the elements of the $U$ matrix are non-zero, the output quantum state $\ket{\psi^{\rm out}}_{c_1d_1}$is entangled one.
Applying a positive operator-valued measure (POVM) $\mathcal{M}=\{M^x\}$ to mode $c_1$ yields a measurement outcome $x_1$ with probability
\begin{equation}
    P(x_1)=\bra{\psi^{\rm out}}M^{x_1}_{c_1}\otimes\mathbb{1}_{d_1}\ket{\psi^{\rm out}},
\end{equation}
and the resulting conditional state of mode $d_1$ is given by
\begin{equation}
    \rho^{\rm out}_{d_1}(x_1)=P(x_1)^{-1}{\rm Tr}_{c_1}\left[M^{x_1}_{c_1}\otimes\mathbb{1}_{d_1}\ket{\psi^{\rm out}}_{c_1d_1}\bra{\psi^{\rm out}}\right],
\end{equation}
where $\mathbb{1}$ is the identity operator, $M^x\geq 0, \sum_x M^x=\mathbb{1}$, and ${\rm Tr}_{c_1}$ denotes a partial trace over $c_1$.
In the case of an ideal photon-number-resolving detector (PNRD), $M^x=\ket{x}\bra{x}$ for $x=0,1,2,\ldots$, and so the resulting state is a pure: $\rho^{\rm out}_{d_1}(x_1)=\ket{\phi^{\rm out}(x_1)}_{d_1}\bra{\phi^{\rm out}(x_1)}$.
The first step $t=1$ ends with transferring the state from the mode $d_1$ to $b_2$:
$\rho^{\rm in}_{b_2}(x_1)=\rho^{\rm out}_{d_1}(x_1)$.

Let $y_\tau:=(x_1,\ldots,x_\tau)$ denote a history of measurement results up to the step $\tau$.
Then we can write down the recursive expressions for all the states that appear at the step $t>1$.
We consider the PNRD case, so all the states are pure.
We also omit explicit labeling of modes to simplify the notations.
The initial state of the modes $a_t,b_t$ is given by
\begin{eqnarray}
\label{start_state}
    \ket{\psi^{\rm in}(y_{t-1})} =\ket{1}\otimes\ket{\phi^{\rm out}(y_{t-1})},
\end{eqnarray} 
where $\ket{\phi^{\rm out}(y_{t-1})}$ comes from the previous step from $d_{t-1}$, and a fresh photon comes in the $a_t$ mode.
The transformation of $\ket{\psi^{\rm in}(y_{t-1})}$ to $\ket{\psi^{\rm out}(y_{t-1})}$ inside the interferometer (from modes $a_t,b_t$ to $c_t,d_t$) is specified by Eq.~\eqref{eq:interferometer}.
The probability to obtain $x_t$ as a measurement outcome, given the history $y_{t-1}$, reads
\begin{equation}
   P(x_t|y_{t-1}) = \bra{\psi^{\rm out}(y_{t-1})}\left(\ket{x_t}\bra{x_t}\otimes\mathbb{1}\right)\ket{\psi^{\rm out}(y_{t-1})},
\end{equation}
while the collapsed state on the mode $d_t$ is given by
\begin{equation}
    \ket{\phi^{\rm out}(y_t)}=\frac{1}{\sqrt{P(x_t|y_{t-1})}} (\bra{x_t}\otimes \mathbb{1})\ket{\psi^{\rm out}(y_{t-1})}.
\end{equation}

The operation of the device corresponds to a kind of quantum-classical stochastic process, where at each step there appear various quantum states ($\ket{\psi^{\rm in}(y_t)}$, $\ket{\psi^{\rm out}(y_t)}$, $\ket{\phi^{\rm out}(y_t)}$), and classical measurement results $x_t$.
We note that the whole process can be considered as a Markovian, since the distribution of states and measurement results at the step $t$ is completely specified by an output state $\ket{\phi^{\rm out}(y_{t-1})}$ at the $(t-1)$-th step.
At the same time, the classical measurement results $x_t$, considered alone, form a non-Markovian random process: the value of $x_t$ is governed by the measured quantum state that depends on the whole history $y_{t-1}$, or more precisely on the sum $\sum_{\tau=1}^{t-1}x_{\tau}$.
In this way, the quantum part provides a Markovian embedding for a random process $(x_t)$.

\section{Reducing uncertainty of quantum measurements}

In contrast to the previous works~\cite{loop_BS_theory1, loop_BS_theory2} on similar loop-based architectures we focus on analyzing entire trajectories in event space rather than a set of individual events sampled at different time steps. The main advantage of this trajectory-based consideration is the possibility to reduce uncertainty of future measurements by utilizing the history of measurement results at previous step, which can be demonstrated by the following example. If $m$ photons came as input on the mode $b$ and one photon on the mode $a$, the interferometer provides transformation of the form
\begin{equation} \label{eq:interferometer-action}
    \ket{1}_a\ket{m}_b \mapsto \sum_{x=0}^{m+1}A_{x,m}\ket{x}_c\ket{m+1-x}_d,
\end{equation}
where
\begin{equation} \label{eq:fock-repr}
    \begin{aligned}
            A_{0,m}&=\sqrt{m+1}u_{12}u_{22}^m,\quad A_{m+1,m}=\sqrt{m+1}u_{11}u_{21}^m,\\
            A_{x,m}&=\sqrt{\frac{m!x(m+1-x)}{(x-1)!(m-x)!}}u_{21}^{x-1}u_{22}^{m-x}\times\\
            &~~~~\left(
                \frac{u_{21}u_{12}}{x} + \frac{u_{22}u_{11}}{m-x+1}
            \right),\quad x=1,\ldots,m.
    \end{aligned}
\end{equation}
After measuring $k$ photons in the mode $c$ (with probability $|A_{k,m}|^2$), the resulting state in $d$ collapses to $\ket{m+1-k}_d$, and so in the next step Eq.~\eqref{eq:fock-repr} can be applied for $m$ replaced by $m+1-k$.
Keeping track of the sum of measured photons $s_{t-1}:=\sum_{\tau=1}^{t-1} x_\tau$ we can redefine the probability of the measurement of $x_t$ as
\begin{equation}
    \Pr[x_t=x|s_{t-1}=s]=|A_{x,t-1-s}|^2.
\end{equation}
This equation has a clear physical interpretation.
During $t-1$ step, preceding the $t$-th one, $t-1$ photons were added to the scheme and $s_{t-1}$ were extracted by the measurement.
So, for the $t$-th step we have 
\begin{eqnarray}
    m=t-1-s_{t-1}
\label{numberofphotons}
\end{eqnarray}
within Eq.~\eqref{eq:interferometer-action}. Put another way, utilizing the history of measurements one chooses a particular trajectory and, therefore, can exclude some events from consideration, which modifies the probability distribution and reduces the uncertainty of measurements.


To quantify correlations induced by a finite measurement history, we introduce the cumulative photon count over a history of depth $k$ preceding the $t$-th step,
\begin{equation}
    s_{t-1}^k = \sum_{i=1}^{k} x_{t-i},
\end{equation}
and define the corresponding event-history correlation function as
\begin{equation} \label{correlation}
    C(k) = \langle s_{t-1}^k x_t \rangle - \langle s_{t-1}^k \rangle \langle x_t \rangle .
\end{equation}

\begin{figure}
  \includegraphics[width=0.93\linewidth]{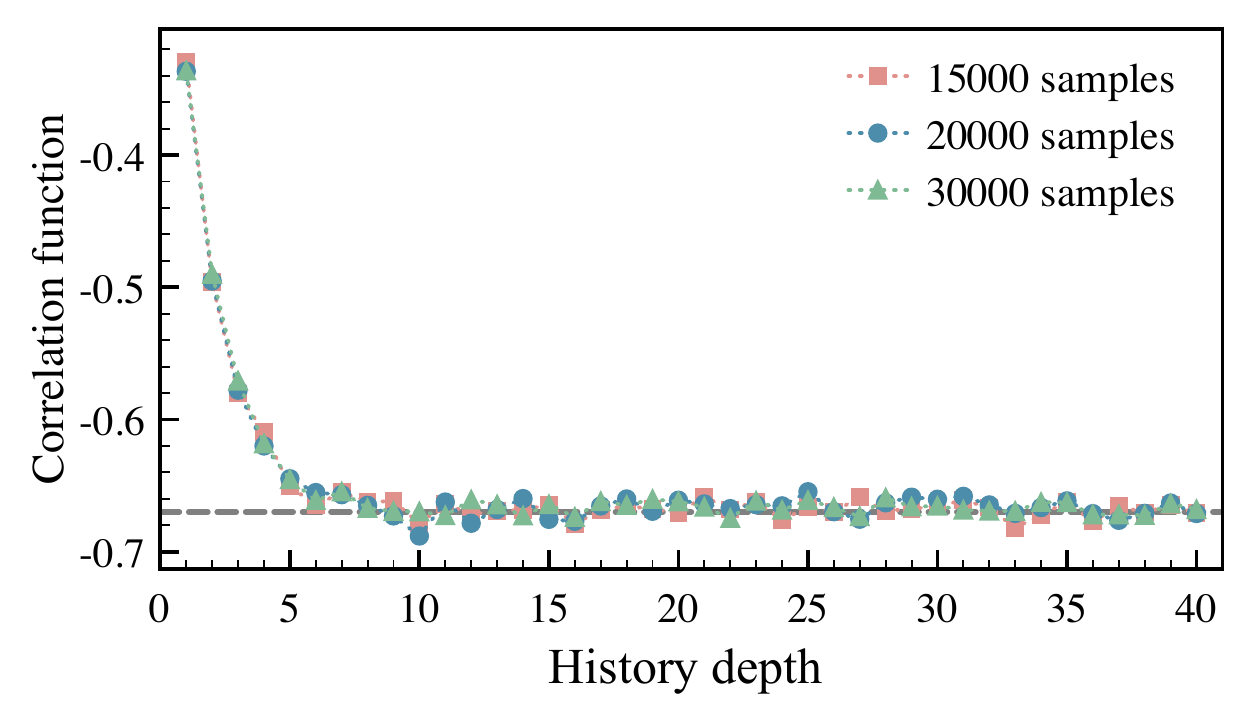}
  \caption{Dependence of the event-history correlation function on the history size. The correlator calculated with Eq.~\eqref{correlation} was approximated by using different sets of 15000, 20000 and 30000 samples. } 
  \label{Finite_depth_corr}
\end{figure}

To estimate this correlation function, we numerically generated a trajectory consisting of $10^5$ events for a 50:50 interferometer
(with $u_{11} = u_{22} = 1/\sqrt{2}$ and $u_{12} = u_{21} = i/\sqrt{2}$)
and sampled trajectory fragments of lengths ranging from 2 to 40 time steps, corresponding to history depths $k=1$ to $k=39$.
For each value of $k$, the number of samples was chosen to be $15\,000$, $20\,000$, or $30\,000$.
The resulting approximate dependence $C(k)$ reveals a pronounced enhancement of temporal correlations in the range $2 < k < 10$
(see Fig.~\ref{Finite_depth_corr}).
For larger values of $k$, the correlation function saturates, indicating the emergence of a stationary correlation regime.

This emergence can be understood analytically.
In Appendix~\ref{app:stationary} we derive the stationary photon-number distribution of the loop mode,
analyze the corresponding transition matrix, and obtain asymptotic expressions for the correlation
function and information-theoretic measures.
Calculations for the 50:50 beam splitter yield $C(1)=-0.333$ and the asymptotic value $C(k\rightarrow\infty)=-0.667$.

As an alternative way to quantify measurement uncertainty, we introduce the following construction.
Let $A$ and $B$ be (generally) dependent random variables.
The standard deviation of $A$ is defined as
$\Delta(A) := \sqrt{\langle A^2 \rangle - \langle A \rangle^2}$,
where $\langle \cdot \rangle$ denotes the expectation value.
We define the conditional standard deviation of $A$ given $B$ as
\begin{equation}
    \Delta(A|B)
    := \sum_b P(b)\sqrt{
        \sum_a a^2 P(a|b)
        - \left(\sum_a a P(a|b)\right)^2
    },
\end{equation}
where $P(b)\equiv \Pr[B=b]$ and $P(a|b)\equiv \Pr[A=a|B=b]$.
Using this definition, we introduce the uncertainty-reduction measure
\begin{equation}
    r(t) := \Delta(x_t) - \Delta(x_t|y_{t-1}).
    \label{uncertainty_decrease}
\end{equation}
Notably, this construction is conceptually analogous to the mutual information discussed in the next section.

\begin{figure}
  \includegraphics[width=0.93\linewidth]{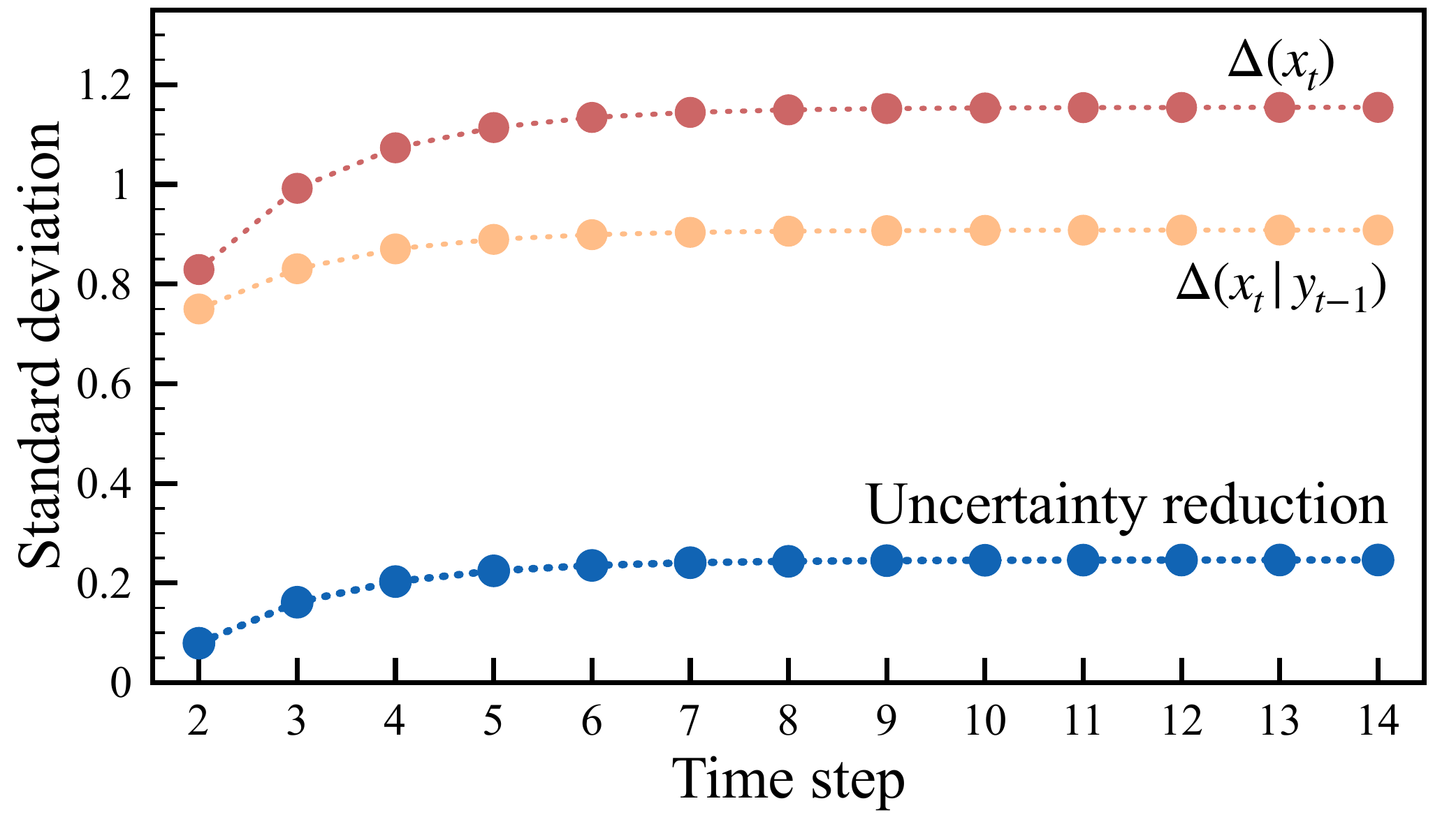}
  \caption{Uncertainty reduction of the quantum measurements calculated with Eq.\ref{uncertainty_decrease}. For these calculations we use the model of the loop-based time-bin interferometer with the 50:50 scattering matrix.}
  \label{Uncertainty}
\end{figure}

In Fig.~\ref{Uncertainty} we show the function $r(t)$ calculated for the loop-based interferometer with the same 50:50 scattering matrix.
The numerical results corroborate our general conclusion that conditioning on the measurement history enables a reduction of uncertainty in subsequent measurements.
Starting from the value $r(2)\approx 0.08$, the uncertainty reduction increases with time and saturates at $r(t)\approx 0.25$ for $t>6$.
This behavior, reminiscent of the relaxation of a dynamical system to a stationary regime, can be traced back to the structure of the underlying probability distributions, as discussed in Appendix~\ref{app:stationary}.

Now we are in a position to discuss quantifying the uncertainty at a given time step when only a finite number of measurements is available. To estimate a physical quantity $A$ from $N$ repeated measurements of a quantum system one can employ different strategies. In the conventional approach, the quantum state is repeatedly prepared, measured, and the obtained outcomes are stored in a classical memory. Repeating this procedure $N$ times yields measurement outcomes $a_1, a_2, \ldots, a_N$, which are then used to estimate the expectation value $\langle A \rangle$. According to the central limit theorem, the standard deviation of the estimator scales as $\Delta(A)/\sqrt{N}$~\cite{Chuang}, where $\Delta(A)$ denotes the standard deviation associated with a single measurement outcome.

Another quantum-enhanced strategy employs entangled quantum states together with collective measurements. In this case the estimation error can scale as $1/N$, corresponding to the so-called Heisenberg scaling~\cite{Giovannetti1}. Since the $1/N$ scaling is commonly regarded as the fundamental limit in standard quantum metrology, further improvements require reducing the effective uncertainty associated with an individual measurement outcome, for instance by exploiting correlations and measurement history~\cite{Giovannetti2}. Using Eq.~\eqref{uncertainty_decrease}, we show that incorporating the measurement history reduces the effective standard deviation $\Delta(A)$.

Potentially, extending the proposed optical reset model with quantum memory elements could provide a platform for studying the interplay between temporal correlations, conditional measurements, and quantum-enhanced metrology. At the same time, most practical applications of reset-like operations in quantum computing~\cite{IPEA1,IPEA2,IPEA3,IPEA4,dissipation1} rely on adaptive single-run protocols, for which classical memory is sufficient.

\section{Information flow}
\subsection{Time-delayed classical mutual information}

Previous studies on quantifying uncertainty in quantum measurements have demonstrated clear advantages of information-theoretic approaches based on Shannon information theory~\cite{Shannon}. For instance, in the entropic formulation of the Heisenberg uncertainty relation~\cite{Deutch,Kraus,Maasen}, Shannon entropy replaces the standard deviation and the corresponding lower bound becomes state independent. Motivated by these ideas, we define the entropic counterpart of the uncertainty reduction introduced in Eq.~\eqref{uncertainty_decrease} as
\begin{eqnarray}
\label{MI_classical}
	J(y_{t-1}: x_{t})=H(x_t) - H(x_t|y_{t-1}).
\end{eqnarray}

Here
\begin{eqnarray}
\label{Shannon}
H(x_t) = - \sum_{x} {\rm Pr}[x_t = x] \log_2 {\rm Pr}[x_t = x],
\end{eqnarray}
is the Shannon entropy associated with the event $x_t$, where $x$ denotes the number of photons detected at the $t$-th time step. This quantity plays the role analogous to the unconditional standard deviation in Eq.~\eqref{uncertainty_decrease} and quantifies the uncertainty of the measurement outcome prior to observing it.

The second term in Eq.~\eqref{MI_classical} is the conditional entropy
\begin{equation}
H(x_t|y_{t-1}) = H(y_t)-H(y_{t-1}),
\end{equation}
where the Shannon entropy of the event history $y_t$ is defined using Eq.~\eqref{Shannon} with ${\rm Pr}[y_t=y]$ replacing ${\rm Pr}[x_t=x]$.

Within information theory, the quantity $J(y_{t-1}: x_t)$ represents the time-delayed mutual information between the event at the current time step, $x_t$, and the measurement history, $y_{t-1}$. In this way, it quantifies temporal correlations accumulated along a given trajectory and provides an information-theoretic measure of the uncertainty reduction analyzed in the previous section.

The mutual information vanishes when the measurement outcome at the $t$-th time step is statistically independent of the previous measurement history. In this case, $H(x_t|y_{t-1}) = H(x_t)$, and, analogously, at the level of the standard deviation one obtains $\Delta(x_t|y_{t-1}) = \Delta(x_t)$,
which corresponds to zero uncertainty reduction, $r=0$.

Finite values of the time-delayed mutual information, Eq.~\eqref{MI_classical}, and weighted uncertainty reduction, Eq.~\eqref{uncertainty_decrease}, indicate the presence of statistical dependence between the event $x_t$ and the trajectory history $y_{t-1}$, such that
\begin{equation}
P(x_t|y_{t-1}) \neq P(x_t).
\end{equation}
As a result, incorporating information about previous measurements makes it possible to reduce the uncertainty of the quantum measurement at the $t$-th step.

Since the mutual information satisfies the bound
\begin{equation}
J(y_{t-1}:x_t)\leq H(x_t),
\end{equation}
its maximal value corresponds to the largest possible reduction of uncertainty associated with the event $x_t$. Therefore, growth (suppression) of the mutual information during time evolution indicates enhancement (reduction) of temporal correlations and, correspondingly, stronger (weaker) predictability of future measurement outcomes.

In the limiting case of the trajectory of minimal duration one can define time-delayed mutual information that describes the information transfer in the system within two consecutive working cycles,
\begin{eqnarray}
\label{eeMI}
	J(x_{t-1}: x_{t}) = H(x_{t}) - H(x_{t}|x_{t-1}),
\end{eqnarray}
where the last term represents the conditional Shannon entropy that describes the events at the $t$-th time step given that the outcome at the previous time ($t-1$) is known, 
\begin{eqnarray}
H(x_{t}|x_{t-1})= - \sum_{xx'} {\rm Pr} [x_t = x, x_{t-1} = x'] \nonumber \\ \times \log_2 {\rm Pr} [x_t = x| x_{t-1} = x'].
\end{eqnarray}
The quantity $J(x_{t-1}:x_t)$ defines the statistical dependence between
outcomes obtained in two consecutive working cycles. Owing to the temporal
ordering of the variables, it can be interpreted as the amount of information
about the future event $x_t$ contained in the previous outcome $x_{t-1}$.
It is worth mentioning that analogs of Eq.~\eqref{eeMI} are used for describing classical dynamical systems of different origin~\cite{Cl_inf_flow1, Cl_inf_flow2, Cl_inf_flow3}.

\begin{figure}[!t]
  \includegraphics[width=0.93\linewidth]{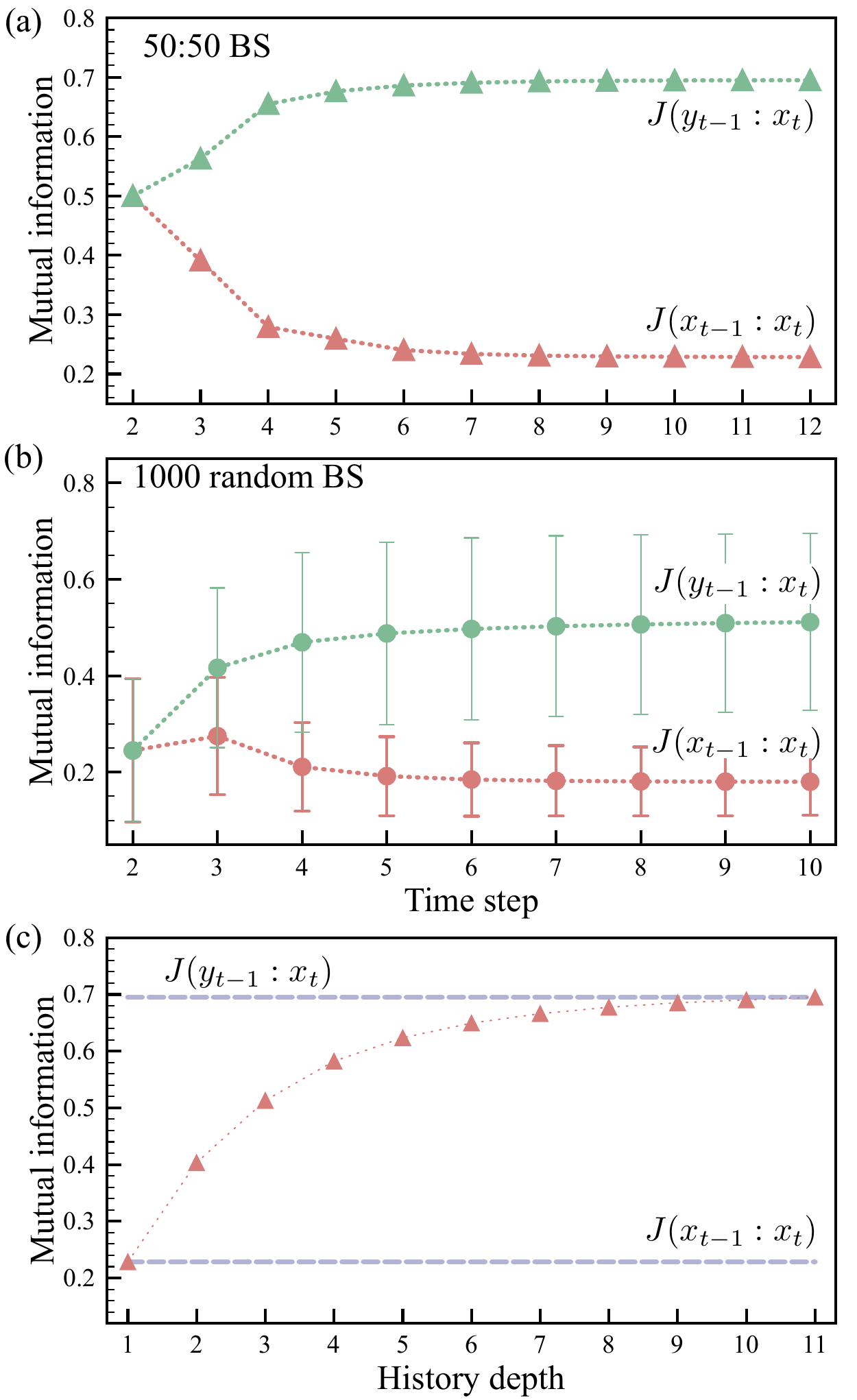}
  \caption{Time-delayed classical mutual information calculated with Eqs.\eqref{eeMI} and \eqref{MI_classical} for the ideal loop-based 50:50 beam splitter (a) and an ensemble of 1000 random interferometers generated with Haar measure (b). (c) Dependence of the event-history mutual information on the history size. In the panel the bottom and top dashed gray lines stand for exact event-event and event-history mutual information, respectively. Red triangles show the values of the event-history approximation, $J((x_{t-k},\ldots x_{t-1}):x_t)$.}
  \label{Class_mutual_info}
\end{figure} 

Having introduced the information-theoretic quantities for characterizing uncertainty of the quantum measurement on the loop-based time-bin interferometer we are going to analyze the numerical results obtained for them with devices characterized by different scattering matrices. The $J(y_{t-1}: x_{t})$ mutual information calculated for the 50:50 (green triangles) and random (green circles) scattering matrices demonstrate the behaviour (Fig.~\ref{Class_mutual_info}) analogous to that obtained for the uncertainty reduction function in Fig.~\ref{Uncertainty}. Namely, there is a fast growth of the mutual information and transition to the stationary regime after the sixth step. A qualitative agreement between time dependencies of $J(y_{t-1}: x_{t})$ and $r(t)$ is expected, since by the definition the mutual information can be considered as an information-theoretic measure of uncertainty reduction about an variable when measuring another one. In our case these variables are the events $x_{t}$ and measurements history $y_{t-1}$. 

It is instructive to analyze the formation of the obtained event-history mutual information time dependence. For that, we first consider the event-event mutual information, Eq.~\eqref{eeMI} calculated for two consecutive time steps. According to Fig.~\ref{Class_mutual_info}(a) $J(x_{t-1}: x_{t})$ for 50:50 BS demonstrates a decreasing trend at starting cycles, it decreases approximately by half in the first steps of evolution and, then, stabilizes at the value of 0.23 for $t>7$. At the same time, the results obtained for an ensemble of the random interferometers [Fig.~\ref{Class_mutual_info}(b)] are characterized by a slight enhancement of the mean mutual entropy at the third time step. Despite difference in statistics at the first time steps both sources demonstrate an equilibration for $t > 6$.    
For the examples we consider (Fig.~\ref{Class_mutual_info}) such an entropy reduction reaches 0.7 and 0.5 bits in the case of the ideal 50:50 and Haar-random interferometers, respectively. In  Appendix~\ref{app:stationary} we provide an interpretation of these results by analyzing the underlying probability functions.  

Formation of the constant gaps between the values of the event-history and event-event types of the mutual information observed at late time steps in Figs.~\ref{Class_mutual_info} (a) and (b) also suggests presence of complex multi-time correlations. Based on this and the calculated event-history correlation function [Eq.~\eqref{correlation}], one can conclude that the two-point event-event correlation functions calculated in Ref.~\cite{loop_BS_theory1}, for which the authors found an exponential decay, do not provide the exhaustive description of the temporal correlations in the system in question. To explore the depth of the temporal correlations of the loop-based interferometer we fix the final time at the value $t=12$ and calculate the classical mutual information of the 50:50 beam splitters varying the history size. From Fig.~\ref{Class_mutual_info} (c) one can see that the $J(y_{t-1}: x_t)$ for $t=12$ can be reliably approximated using $J((x_{t-k},\ldots x_{t-1}):x_t)$ with $k > 8$.   

These findings concerning a finite depth of the time-delayed classical mutual information can contribute to the theory of quantum Markov order developed in Refs.~\cite{Modi1,Modi2}. Based on partitioning time steps into blocks of history, memory and future, such a theory demonstrates that any non-Markovian quantum process has infinite Markov order with respect to a generic instrument sequence used for measurements. However, it was shown that finite memory effects can be revealed for specific choices of instruments. Here, we propose the concrete optical platform that features a distinct PNRD instrument and allows to test the Markov order theory in real experiments \cite{Exp4, Exp5, Exp6}. From the theoretical side, taking into account the estimated depth of the temporal correlations [Fig.\ref{Class_mutual_info} (c)] one should consider increasing the total number of time steps in our exact simulations of the optical reset to accurately reproduce the time structure -- consisting of the history, memory and future blocks -- implied by quantum Markov order theory. We leave this for a future investigation.

\subsection{Time-delayed classical--quantum mutual information}

To conclude this section, we characterize the information flow at the level of the quantum loop subsystem, represented by modes $b$ and $d$. For this purpose, we introduce the classical--quantum mutual information between the measurement history $y_t$ and the conditional output state $\ket{\phi^{\rm out}(y_t)}$ in mode $d$ at the time step $t$:
\begin{eqnarray}
\label{quan_cl_mutual}
   \mathcal{I}(y_t:\phi^{\rm out}(y_t)) =
   S\left(\sum_y \Pr[y_t=y] \sigma(y)\right) \nonumber \\
   - \sum_y \Pr[y_t=y] S\left(\sigma(y)\right),
\end{eqnarray}
where $\sigma(y):=\ket{\phi^{\rm out}(y)}\bra{\phi^{\rm out}(y)}$
is the conditional state of the loop subsystem associated with the history $y$, and
$S(\sigma)=-{\rm Tr}(\sigma\log_2\sigma)$ is the von Neumann entropy. 
Put another way, the quantity $\mathcal{I}$ is the Holevo information of the
ensemble of conditional loop states generated by different measurement
histories. It quantifies how much classical information about the realized
history $y_t$ is encoded in the corresponding conditional quantum state of
the loop subsystem. We emphasize that $y_t$ includes the measurement outcome
at the $t$-th step, so that $\ket{\phi^{\rm out}(y_t)}$ is conditioned on the
full trajectory up to and including this step.

For an ideal PNRD measurement, each conditional state $\sigma(y)$ is pure, and
therefore the second term in Eq.~\eqref{quan_cl_mutual} vanishes. The first
term is the entropy of the ensemble-averaged loop state,
\begin{equation}
    \bar{\rho}_d(t)=\sum_y \Pr[y_t=y]\sigma(y).
\end{equation}
In the numerical simulations, we observe a thermalization-like relaxation of
this averaged state: after a transient time, $\bar{\rho}_d(t)$ approaches a
stationary state $\rho_d^{\rm stat}$. Consequently, in this regime the
classical--quantum mutual information approaches
\begin{equation}
    \mathcal{I}(y_t:\phi^{\rm out}(y_t)) \to S(\rho_d^{\rm stat}).
\end{equation}
As discussed in Sec.~\ref{sec:thermalization}, this stationary state can be
close to a thermal state, which provides an additional interpretation of the
saturation of the information flow in the loop subsystem.

\section{Reconstructing history}

\begin{figure}[!t]
  \includegraphics[width=0.99\linewidth]{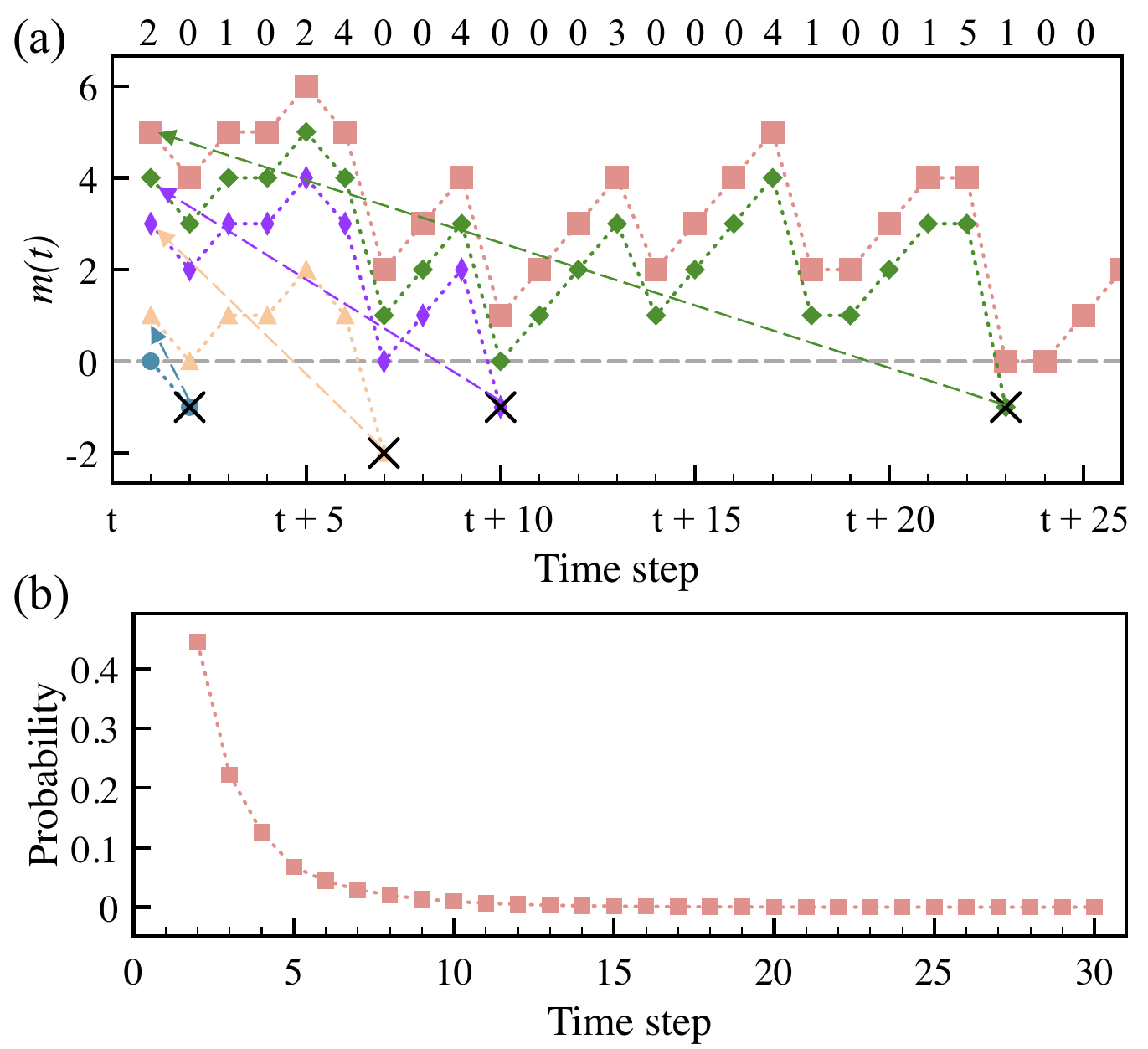}
  \caption{(a) Schematic representation of the proposed approach to restore the number of photons inside the interferometer. Numbers above the figure frame correspond to 25-step trajectory fragment used as an example. Each time $m(t)$ is negative one sets it to zero, updates all previous time steps and restarts the procedure. The exact values are represented by the red squares. The corresponding outputs on the mode $c$ are given above the graph. (b) Probability of the last negative $m(t)$ occurring at each specific time step of the randomly selected sequence of measurements.}
  \label{fig:deltas}
\end{figure}

Having the history of measurements ($y_{t-1}$), the total number of photons inside the interferometer at the $t$-th time step can be calculated exactly and, as discussed in the methodological part, is equal to $m(t) = t-1-\sum_{\tau=1}^{t-1} x_\tau$. This allows one to reduce uncertainty of measurement at the given time step, which can be quantified with the event-history mutual information and correlation function [Eqs.~\eqref{MI_classical} and \eqref{correlation}]. In this section, we explore the possibility to restore the $m (t)$ function from scratch for any trajectory fragment starting at arbitrary time step without additional information.    

The procedure we propose for reconstructing the number of photons in the loop can be explained by using a particular example of trajectory fragment obtained for the 50:50 loop beam splitter and presented in the top of Fig.\ref{fig:deltas}(a). We assume that there is no information about actual starting and ending times when this fragment was obtained. Thus, it is natural to set the initial values of $m(t)$ function to zero. In the given example the first measurement reveals 2 photons in the mode $c$, ($x_{1} = 2$) which according to Eq.~\eqref{numberofphotons} results in negative $m(t=2)$ at the second time step of the given fragment (blue circles). To avoid this the initial value of the $m(t)$ function should be corrected by adding $x_{1}-1$, so that $m(t=1) = 1$ and $m(t>1) = 0$. Then we restart the procedure and calculate the $m(t)$ function till the 7th time step (orange triangles). Having $m(6)=1$ and $x_6=4$ we again end up with the negative $m(7)=-2$ which means that we should add another two photons to the initial value. After repeating this procedure several times, one can fully recover the $m(t)$ function at each time step (red squares).   

Figure~\ref{fig:deltas}b shows the probability that the last negative event will occur at a specific time step calculated using the $10^7$ random fragments of the whole trajectory. The resulting value of $\Pr[m(t) < 0]$ saturates below 1\% at $t=10$, which correlates with the behavior of $C (k)$ and $J(y_{t-1}: x_{t})$. The last possible time step when a negative event occurs is $t=30$, which can be used as an upper bound for measurements count when reconstructing the function $m(t)$.

\section{Applications}
In this section we discuss potential applications of the optical reset introduced in the present work. Such a discussion also reveals unexplored features of the model.

\subsection{Extending boson sampling to the temporal domain}

Loop-based time-bin interferometers provide a promising platform for
boson sampling experiments~\cite{Exp4,Exp5,Exp6}. In the standard
setting, boson sampling aims at generating samples from a probability
distribution over an exponentially large output space produced by a
linear-optical interferometer, which is believed to be hard to simulate
classically. Achieving this regime typically requires the number of
output modes to scale at least quadratically with the number of input
photons, $N^{\rm out}_{\rm modes} \gtrsim (N^{\rm in}_{\rm ph})^2$,
in order to remain in the collision-free regime relevant for complexity
arguments~\cite{Aaronson1}. However, such spatial scaling quickly becomes
a major experimental bottleneck.

Time-bin loop architectures offer an alternative by enlarging the
effective sampling space through temporal multiplexing, reusing the
same optical hardware over multiple cycles. A time-bin boson sampler
with $L$ photons circulating in the loop, injecting $N^{\rm in}_{\rm ph}$
photons per cycle over $t$ steps, can be viewed as an effective device
with $tN^{\rm in}_{\rm ph}$ photons and
$(N^{\rm out}_{\rm modes}-L)t+L$ output modes~\cite{Exp6}. In this way,
the size of the sampling space grows not only with the number of spatial
modes, but also with the temporal depth of the experiment.

The presence of temporal correlations and classical memory leads to two
distinct operational regimes. When the memory depth exceeds the
correlation length, past measurement outcomes carry significant
information about future events, reducing the effective uncertainty of
the sampling process. This corresponds to a memory-assisted regime,
where classical side information introduces partial predictability in
the output statistics. In contrast, when sampling is performed with
delays exceeding the correlation depth, temporal correlations become
negligible, and the process approaches a memoryless limit with
approximately independent outcomes, resembling conventional boson
sampling with an enlarged effective size. The ability to interpolate between these regimes by controlling the use
of classical memory provides additional flexibility for designing
computational architectures based on time-multiplexed boson sampling.

This transition from a temporally correlated, measurement-driven
dynamics to an effectively memoryless regime is conceptually related to
fully coherent approaches such as Quantum Signal Processing (QSP)
\cite{QSP1,QSP2}, where conditional logic is embedded directly into a
larger unitary evolution. While such coherent protocols reduce the need
for intermediate measurements, their implementation in photonic
platforms typically requires unfolding temporal dynamics into spatially
extended interferometers, leading to increased losses. In contrast, the
loop-based reset model exploits time-bin multiplexing and intermediate
measurements, enabling repeated use of the same hardware and trading
coherence for reduced experimental complexity. Moreover, unlike QSP
designed for static quantum registers, this architecture naturally
supports continuous photon streams and dynamical state manipulation over
extended time scales.

\subsection{Exploring the quantum thermalization process} 
\label{sec:thermalization}
The interplay between information processing and thermodynamics is
central to Landauer's principle, which relates entropy reduction to
heat dissipation~\cite{Landauer}. Its quantum formulation was developed
by Reeb and Wolf~\cite{Reeb}, who considered a system interacting with
a thermal environment via a joint unitary evolution.

\begin{figure}
  \includegraphics[width=0.93\linewidth]{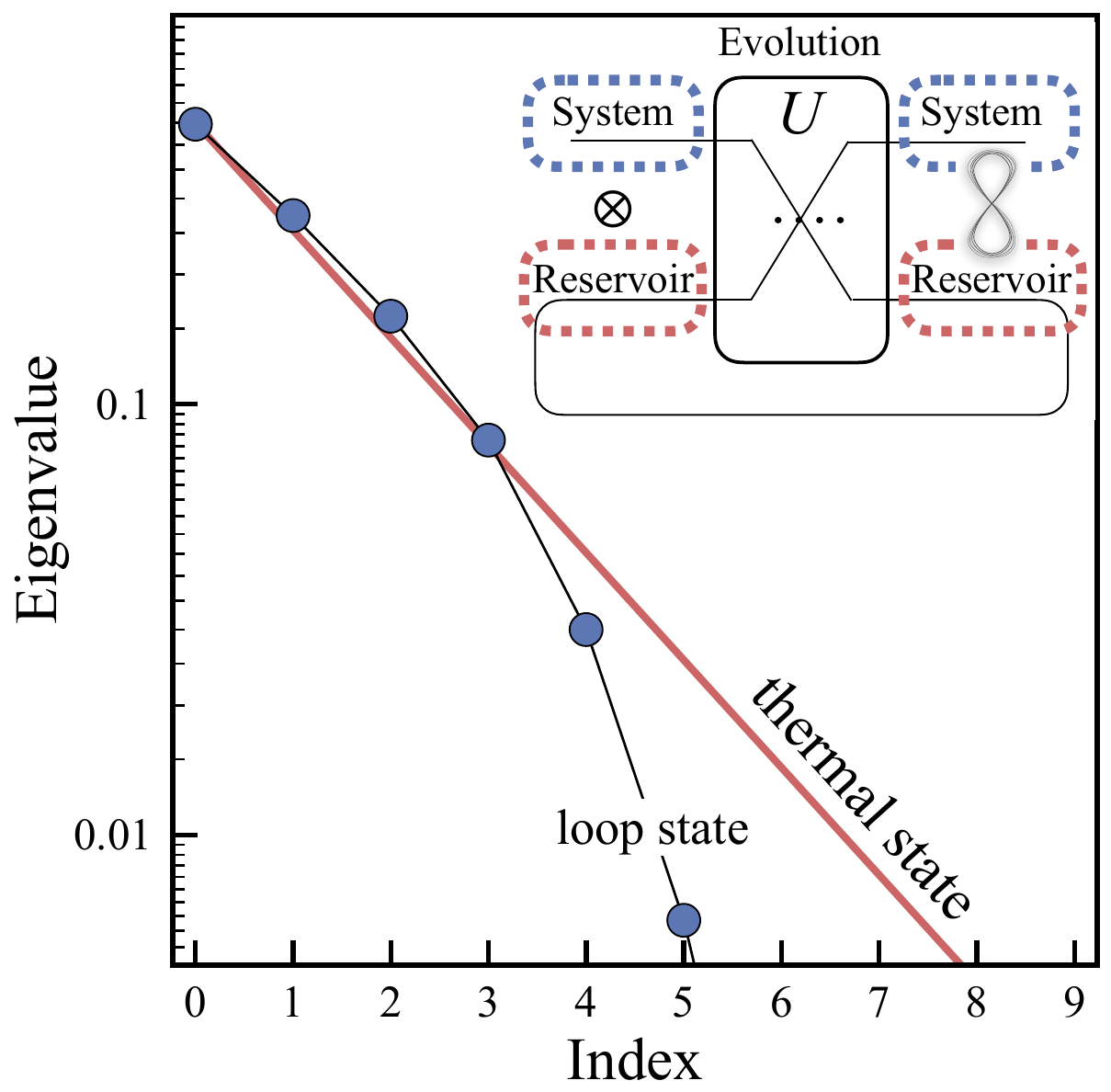}
  \caption{Ordered eigenvalues of the mixed state of the loop subsystem. The data were obtained with the 50:50 scattering matrix. The von Neumann entropy of the mixed state in the $d$ mode is equal to 1.95. Inset gives the scheme of the time-bin interferometer without detector, which enables the formation of the mixed quantum state in the loop subsystem in 12 time steps. Red line denotes the exponential fit, $E_{n} = 0.4497\times 10^{-0.2494n}$ for the first 4 points of the data.}
  \label{qLandauer}
\end{figure}

In our setup, this framework can be naturally adapted by identifying
the modes $a$ and $c$ as an effective ``system'', while the loop modes
$b$ and $d$ act as an effective ``reservoir'' (Fig.\ref{qLandauer}). Each cycle starts from
an uncorrelated state [Eq.~\eqref{start_state}], followed by a unitary
evolution that entangles these subsystems.

A mixed state with thermal-like properties can be generated in the loop
by suppressing intermediate measurements. As shown in
Fig.~\ref{qLandauer}, the eigenvalue spectrum of the reduced density
matrix of the loop subsystem exhibits an approximately exponential
profile on a logarithmic scale, consistent with a thermal distribution.

Adapting the result of Ref.~\cite{Reeb} to this setting, the average heat
exchange can be written as
\begin{equation}
\label{quantum_Landauer}
\beta \Delta Q = \Delta S + \mathcal I (c : d)
+ \mathcal D (\rho_{d} \Vert \rho_{b}),
\end{equation}
where $\beta$ is an effective dimensionless inverse temperature, $\Delta Q = \langle n_d \rangle - \langle n_b \rangle$ denotes the
change in the photon number of the reservoir, $\Delta S = S(\rho_a) -
S(\rho_c)$ is the entropy change of the system, $\mathcal I(c:d)$ is the quantum mutual information between modes $c$ and $d$, characterizing correlations generated by the joint unitary
evolution, and $\mathcal D$ is the relative entropy
between input and output reservoir states.

In the stationary regime, the loop subsystem approaches a fixed-point
distribution with $\rho_b \approx \rho_d$. In this limit, the relative
entropy term vanishes and the net heat flow becomes negligible,
$\Delta Q \approx 0$, while the entropy change of the system is
compensated by system–environment correlations. This correlation-driven
balance represents a distinct non-equilibrium steady state that may be
accessible experimentally.

To probe non-trivial thermodynamic behavior, one can perturb this
steady-state by varying the interferometer parameters after the system
has equilibrated, thereby inducing transient heat flows and deviations
from the stationary regime.

\subsection{Other potential applications}
Our discussion of possible applications of the proposed optical reset model naturally concludes with examples already demonstrated on superconducting qubit platforms. As mentioned in the introduction, reset operations can significantly enhance the capabilities of quantum computing architectures and have been employed in iterative phase estimation \cite{IPEA1, IPEA2, IPEA3, IPEA4}, quantum autoencoders \cite{QAE1,QAE2}, engineered dissipation \cite{dissipation1}, and quantum error correction~\cite{QEC1, QEC2}. In principle, these protocols could also be implemented on optical platforms using dual-rail encoding~\cite{dual-rail}, where a qubit is represented by a superposition of single-photon states distributed over two spatial modes. Within this framework, single-qubit gates are realized with beam splitters and phase shifters, whereas two-qubit entangling operations such as the CNOT gate require postselection over auxiliary modes. Since existing software platforms for photonic quantum simulations, such as Perceval~\cite{perceval}, do not currently support reset operations or measurement-conditioned dynamical circuits, extending them in this direction represents an important avenue for the future research.

\section{Conclusion and outlook}

To sum up, we have developed a theoretical model of the quantum optical $\sf reset$ on the basis of the existing interferometers with loop-based architecture \cite{Exp4, Exp5, Exp6} that allows quantum interference between photons loaded into the device at different times. Our idea is based on using photon-number-resolving detectors and classical memory that operate synchronously with each other. These components enable decreasing uncertainty of the quantum measurement of the $\sf reset$ operation at each time step with the simplest computation (summation) on the classical hardware. In this way one gets a distinct mean to control conditional wave functions generated with quantum device, which paves the way to realize the idea on dynamical quantum computing \cite{IPEA4} with optical devices.

In our study we fully concentrate on the analysis of the loop-based architecture that provides quantum interference between the nearest time steps. However, as it was argued in Refs.\cite{Exp1, Exp5} more complex temporal connections can be potentially realized in such experiments. For instance, by manipulating time delay of the loop part of the quantum device one can create distinct random-like or ordered patterns of temporal correlations which can be still probed and characterized with the non-local-in-time measures developed by us in this work. 

Our proposed optical reset model can be employed as a tool to probe and quantify experimental defects. By estimating the average number of output photons in the mode $c$, one can monitor the system. In the ideal stationary state, this value should be equal to 1. Any observed deviation provides a direct metric to quantify the collective impact of experimental imperfections (Appendix B). While these defects can originate from the source, the optical circuit, or the detector, a comprehensive decomposition of these noise sources constitutes a significant separate study, which we leave to future work. 

Another important feature of our protocol is that it can be interpreted as a quantum walk in the photon-number space. In this analogy, the hidden loop acts as the walker, while the mode $c$ serves as an infinite-dimensional quantum coin. Unlike standard quantum walks that utilize finite-dimensional coins (e.g., qubits or polarized photons) \cite{quantum_walk1, quantum_walk2}, our framework exploits the infinite-dimensional Fock space for both parties. This allows for a more complex redistribution of photon statistics, which we characterize through the steady-state distribution. In turn, by adjusting measurement frequency the system can exhibit features similar to the Quantum Zeno effect~\cite{Zeno1, Zeno2, Zeno3}, where the state of the hidden loop is stabilized by the act of observation. This repeated projection prevents stabilizing system into a mixed state. Importantly, without measurements on the mode $c$ the system likewise reaches a stationary mixed state due to its intrinsic feedback structure. The properties of such a mixed state is similar to the thermal ones as we describe in the section devoted to the applications.

\section*{Acknowledgments}
We thank Roman Morozov for fruitful discussions.  

\appendix
\section{Events probabilities and an example of uncertainty decrease} \label{app:example}
By the construction, the quantum loop-based interferometer (Fig.~\ref{loop_BS}) generates classical data which are event series of photon numbers [for example, $(0,0,1,2,\ldots)$, $(1,0,2,1,\ldots)$, and so on] detected at different time steps over individual runs. Importantly, the dictionary of events, which can be generated, expands at each new time step, which requires estimation of classical resources for storing the generated data in classical memory. This can be done with the information theory developed by Shannon~\cite{Shannon}, where the basic quantity, which allows to define different information measures, is a probability distribution of a random variable. In our case, we consider the probabilities of two types. ${\rm Pr}(y_{t})$ is the probability to obtain a sequence of events (event series) during a run of $t$ time steps. Another one is the probability to generate an event at the given time step, ${\rm Pr}(x_{t})$. Both functions are controlled by the scattering matrix $U$ and input quantum state in the mode $a$.

\begin{figure}[!b]
  \centering
  \includegraphics[width=0.93\linewidth]{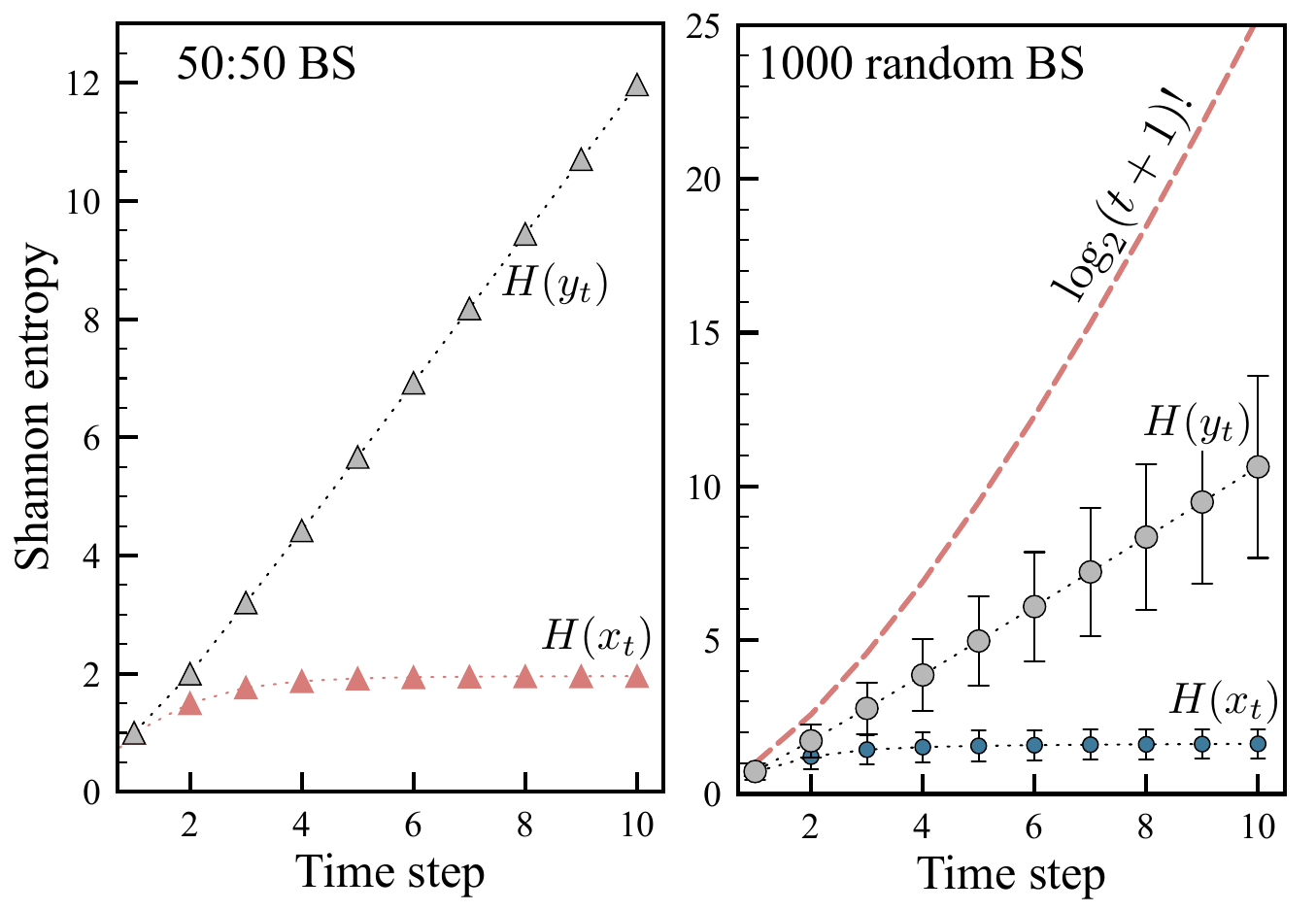}
  \caption{Comparison of the Shannon entropy calculated with probabilities of the events detected in the output mode $c$ at the given time step, $H(x_t)$ and those for the event series, $H(y_{t})$. The results obtained for the ideal 50:50 beam splitter (left panel) and the data averaged over 1000 independent beam splitters with random scattering matrices generated using Haar measure (right panel). Red dashed line denotes the limiting Shannon entropy calculated for the series of the independent equally probable events.}
  \label{Shannon_entropy}
\end{figure} 

Figure~\ref{Shannon_entropy} demonstrates the comparison of $H(x_{t})$ and $H(y_{t})$ calculated for the ideal 50:50 beam splitter (left panel) and the ensemble of 1000 interferometers characterized by random scattering matrices $U$ generated with the Haar measure (right panel). The results were obtained with the Perceval package~\cite{perceval}. In both cases we observe a clear difference in the behavior of data describing event series and those corresponding to an individual event. While the former demonstrates a linear growth, the latter features a saturation at the sixth time step. For the 50:50 beam splitter, the saturation value of the Shannon entropy is equal to 1.96 bits (see Appendix~\ref{app:stationary} for the derivation of this value), which is larger than the mean $H(x_{t})$ of approximately 1.6 bits in the case of the ensemble of 1000 random beam splitters. Taking into account the standard deviation of about 0.48 bits in the worst case, about 2.1 bits per event are required to encode the outcome at any time step.
Thus, the simplest loop-based interferometer (Fig.~\ref{loop_BS}) reveals two distinct dynamical regimes with the critical time $t_c \approx 6$. This agrees with the results reported in Ref.~\cite{loop_BS_theory1} where the authors analyze the entanglement entropy of the loop mode and conclude on its saturation when the rate of photons going into the system is equal to the rate with which photons leak out to the detection mode. 
This value also agrees with the characteristic saturation time for the 50:50 beam splitter $\tau_{\rm sat}=1.4$ calculated according to Appendix~\ref{app:stationary}.

\begin{figure}
  \centering
  \includegraphics[width=0.95\linewidth]{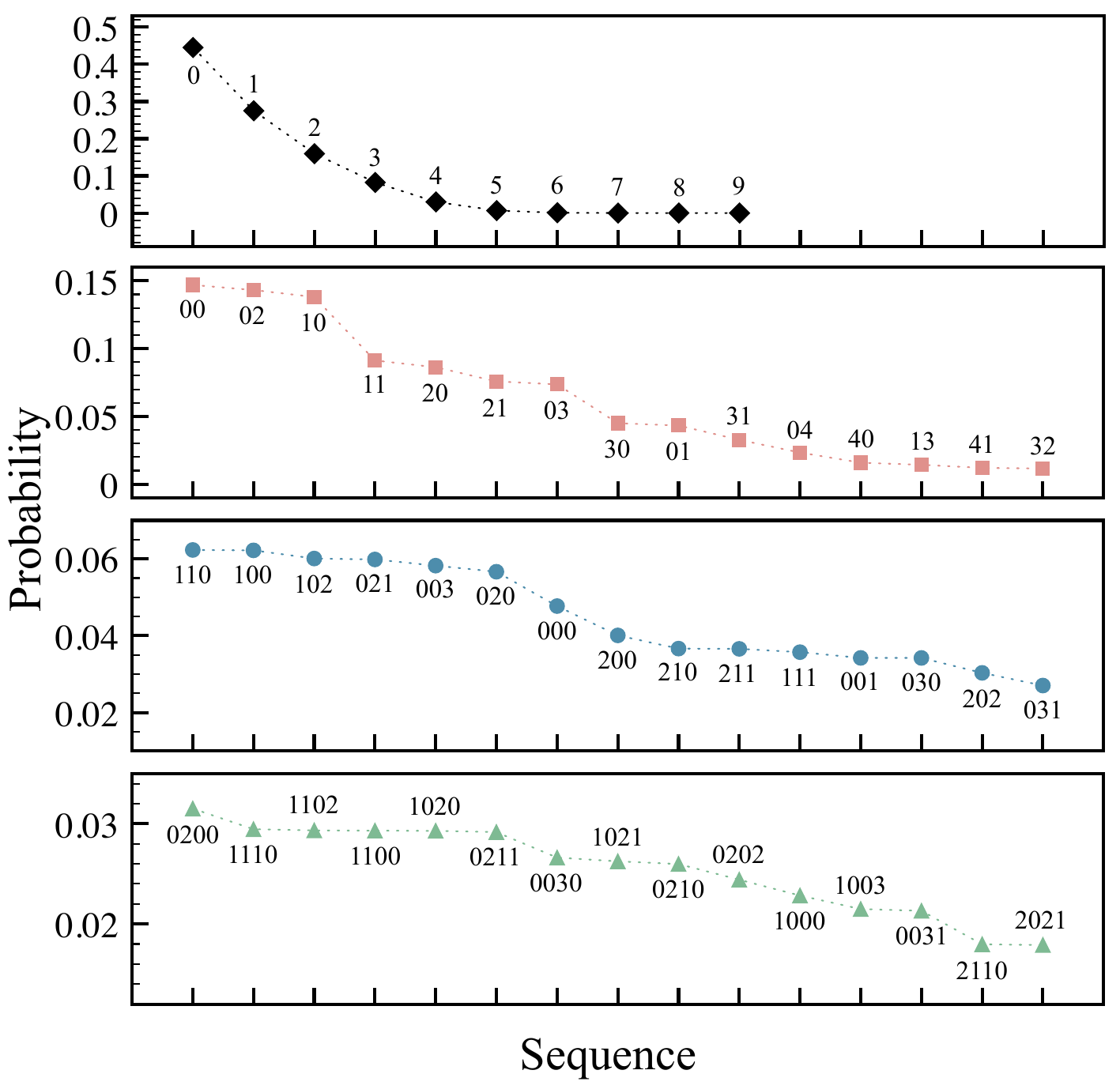}
  \caption{Most probable sequences of events of different length from 1 to 4. Probabilities are estimated for a $10^6$-step trajectory obtained at the output mode of the ideal 50:50 boson sampler samples.}
  \label{sequences_5050}
\end{figure}

The observed difference between $H(x_{t})$ and $H(y_{t})$ can be explained with the probability distributions estimated for individual events and event trajectories of different length. The results in Fig. \ref{sequences_5050} show that for a 50:50 BS the largest contributions come from events with zero or only a few photons. These probabilities are 0.45, 0.27, 0.16, 0.08, and 0.03 for photon numbers from 0 to 4 respectively, while for $x >5$ the values become negligible. The situation changes when considering event series because increasing the trajectory length leads to a greater number of sequences with approximately the same probability.  

It is also instructive to estimate the Shannon entropy for trajectories of different length in the limiting instance, when all events at a given time step are equally probable and they are independent from the previous measurements in the same run. In this limiting case the Shannon entropy of the event trajectory is defined as $\log_2[ (t+1)!]\approx (t+1)\log_2 (t+1)- (t+1)\log_2e$ that grows much faster (the red dashed line in Fig.~\ref{Shannon_entropy} right) than the dependencies obtained for BS samplers we consider in this work. This confirms existing imbalance between probabilities of different trajectories that is due to the loop-based structure of the device we consider. In Appendix~\ref{app:memory} we estimate classical memory resources needed to store the generated trajectories.

\begin{figure}
  \centering
  \includegraphics[width=0.85\linewidth]{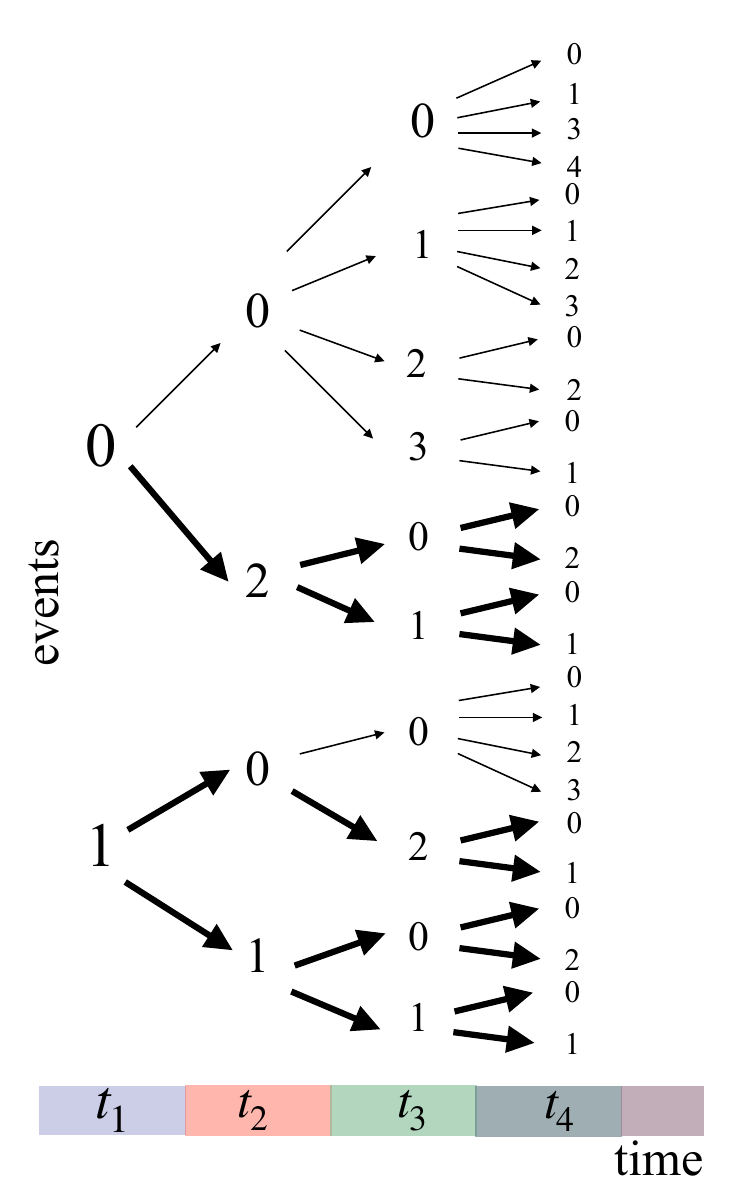}
  \caption{Example of the 4-time run trajectories in the event space obtained for the ideal 50:50 boson sampler. Bold arrows denote the event series of the maximal probability.}
  \label{event_time_5050}
\end{figure}

Figure~\ref{event_time_5050} illustrates the idea of reducing the uncertainty with counting the number of photons on the mode $c$. It gives all the 4-step event series obtained for the ideal 50:50 beam splitter with $u_{11} = u_{22} = \frac{1}{\sqrt{2}}$ and $u_{12} = u_{21} = \frac{i}{\sqrt{2}}$. Without an additional information, there is non-zero probability to detect '0', '1', '2', '3' or '4' event in the output mode $c_{t}$ at the 4th time step.  However, tracking a specific measurement history up to the third time step, for instance, $y(t_3) = (1,0,2)$ allows one to considerably reduce the number of possible alternatives and can expect to detect either `0' or `1' events at the fourth step.

\section{Stationary regime} \label{app:stationary}

Here we derive the information-theoretic measures and correlation functions for the stationary regime that emerges at large time steps ($t \gg 1$). 
To this end, we first introduce the transition matrix $\Pi(m|m')$, which defines the conditional probability of having $m$ photons in mode $b_{t+1}$, given $m'$ photons were present in mode $b_t$.
This probability is given by the squared probability amplitude,
\begin{equation}
    \Pi(m|m') := |A_{m'+1-m,m'}|^2,
\end{equation}
corresponding to the detecting of $x = m' + 1 - m$ photons in the output mode $c_t$ from an input of $m'$ photons in $b_t$ [see Eq.~\eqref{eq:fock-repr} for the explicit form of the matrix element $A_{x,n}$ for a beam splitter described by matrix $U$]. 
Consequently, the probability of detecting exactly $x$ photons in mode $c_t$ is
\begin{equation}
    P(x|m') = |A_{x,m'}|^2.
\end{equation}

The functions $\Pi(m|m')$ and $P(x|m')$ constitute stochastic matrices, whose columns sum to unity.
The one-step update of the photon number distribution $\pi_{t}(m)$ in mode $b_t$ is governed by the Bayesian  rule:
\begin{equation}
    \pi_{t+1}(m) = \sum_{m'}\Pi(m|m')\pi_t(m').
\end{equation}
The stationary distribution $\pi_{\rm st}(m)$ is the eigenvector of this transition matrix corresponding to a unit eigenvalue, satisfying
\begin{equation}
    \pi_{\rm st}(m) = \sum_{m'}\Pi(m|m')\pi_{\rm st}(m').
\end{equation}
Consequently, for $t \gg 1$, the distribution of the number of detected photons $x_t$ is given by marginalizing over this stationary state:
\begin{equation}
    \Pr[x_{t} = x] \simeq \sum_{m=0}^\infty P(x|m)\pi_{\rm st}(m) =: p_{\rm st}(x).
\end{equation}
These stationary distributions, $\pi_{\rm st}(m)$ and $p_{\rm st}(x)$, form the basis for calculating information-theoretic measures and correlation functions.
We also note that a characteristic saturation time (measured in the number of steps) can be defined as
\begin{equation}
    \tau_{\rm sat} = -\frac{1}{\log \lambda_{\rm max}},
\end{equation}
where $\lambda_{\rm max}$ denotes the largest eigenvalue of the transition matrix $\Pi(m|m')$ different from unity.

First, we obtain a stationary value of Shannon entropy for the photon counts:
\begin{equation}
    H(x_{t}) \simeq  H[ p_{\rm st} ],
\end{equation}
where $H[p]$ denotes Shannon entropy of a probability distribution $p$.
The conditional entropy $H(x_t|y_{t-1})$ in the limit of large $t$ can be expressed as the entropy of the detected photon number conditioned on the input photon number:
\begin{equation}
    H(x_t|y_{t-1}) \simeq  \sum_{m=0}^\infty \pi_{\rm st}(m) H[P(\cdot|m)],
\end{equation}
where $P(\cdot|m)$ denotes $m$-th column of the stochastic matrix.
The asymptotic mutual information $J(y_{t-1}:x_t)$ is given by the difference between the stationary entropy $H(x_t)$ and the stationary conditional entropy $H(x_t|y_{t-1})$.

To analyze the correlation properties between photon counts at two consecutive steps, $t$ and $t+1$, we consider their joint probability distribution in the stationary regime ($t \gg 1$):
\begin{multline}
P[x_t=\alpha, x_{t+1}=\beta] \\
\simeq \sum_{m=0}^{\infty} \pi_{\rm st}(m) P(\alpha|m) P(\beta|m-\alpha+1).
\end{multline}
Here, each term of the sum corresponds to the following sequence of events: we begin with $m$ input photons in mode $b_t$ (and one photon in $a_t$), detect $\alpha$ photons in mode $c_t$, and consequently obtain $m-\alpha+1$ photons in modes $d_t$ and $b_{t+1}$.

The stationary value of the correlation function~\eqref{correlation} for large $t$ and $k$ ($t\gg k\gg 1$) can be derived as follows.
If $k$ significantly exceeds the characteristic thermalization time, the photon numbers $m_{t-1-k}=:m$ and $m_{t-1}=:m'$ in modes $b_{t-1-k}$ and $b_{t-1}$, respectively, can be treated as independent random variables distributed according to $\pi_{\rm st}$.

The total number of detected photons within the interval of steps $t-1-k,\ldots,t-1$ is then given by
\begin{equation}
    N^k:=m+k-m'.    
\end{equation}
Hence, the correlation function takes the form
\begin{equation}
    C(k)=\sum_{m,m',x} xN^k\pi_{\rm st}(m)\pi_{\rm st}(m') P(x|m')-k,
\end{equation}
which can be simplified to
\begin{multline}
    C(k)=\sum_{m,m',x} x(m-m')\pi_{\rm st}
    (m)\pi_{\rm st}(m') P(x|m'),
\end{multline}
showing that $C(k)$ is independent of $k$ in the stationary regime.
Here we have used the fact that the mean number of detected photons over $k$ steps equals $k$, since the average detection rate is one photon per step.

\begin{figure}[!t]
  \centering
  \includegraphics[width=0.93\linewidth]{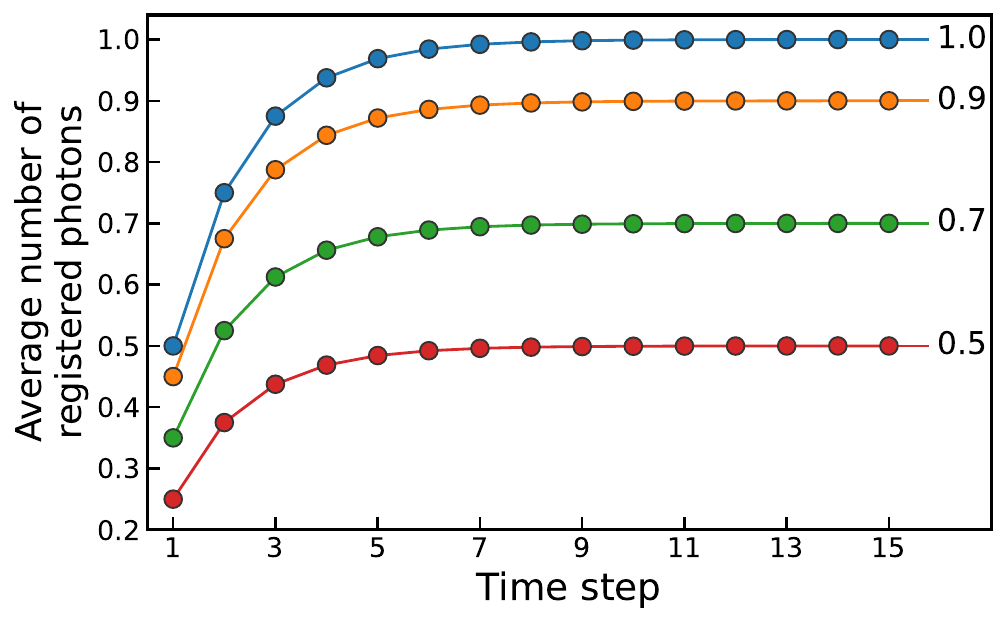}
  \caption{Time dependence of the average number of photons registered in mode $c$ for different detector efficiencies $\eta$, indicated on the right. These calculations were performed with the 50:50 scattering matrix.}
  \label{losses}
\end{figure}

The stationary regime described above is a key feature of our protocol.
For an ideal detector, the average number of photons registered in the output
mode $c$ approaches unity, providing a natural reference value for monitoring
experimental imperfections. To illustrate the effect of finite detection
efficiency, we model the photon-number-resolving detector by assuming that
each photon arriving in mode $c$ is registered independently with probability
$\eta$. Thus, if $x$ photons reach the detector, the registered photon number
$k$ follows the binomial distribution
\begin{equation}
    \Pr(k|x)
    =
    \binom{x}{k}\eta^k(1-\eta)^{x-k}.
\end{equation}
Accordingly, the average registered photon number is related to the actual
average photon number in mode $c$ by
\begin{equation}
    \langle k_t\rangle
    =
    \eta \langle x_t\rangle.
\end{equation}
Figure~\ref{losses} shows the resulting time dependence for several values of
the detector efficiency. Since the ideal output flux approaches
$\langle x_t\rangle=1$ in the stationary regime, the average registered photon
number approaches $\eta$. The detector inefficiency therefore rescales the
observed photon number while leaving the underlying evolution of the optical
loop unchanged. In particular, photons that are not registered have already
left the interferometer through mode $c$ and do not remain in the loop.
Nevertheless, finite detector efficiency produces a mismatch between the
actual photon number emitted from the device and the corresponding classical
measurement record. Such a mismatch may affect protocols that reconstruct the
loop occupation or condition subsequent operations on the detected
measurement history.

\section{Classical resources} \label{app:memory}

\begin{figure}[!t]
  \centering
  \includegraphics[width=0.93\linewidth]{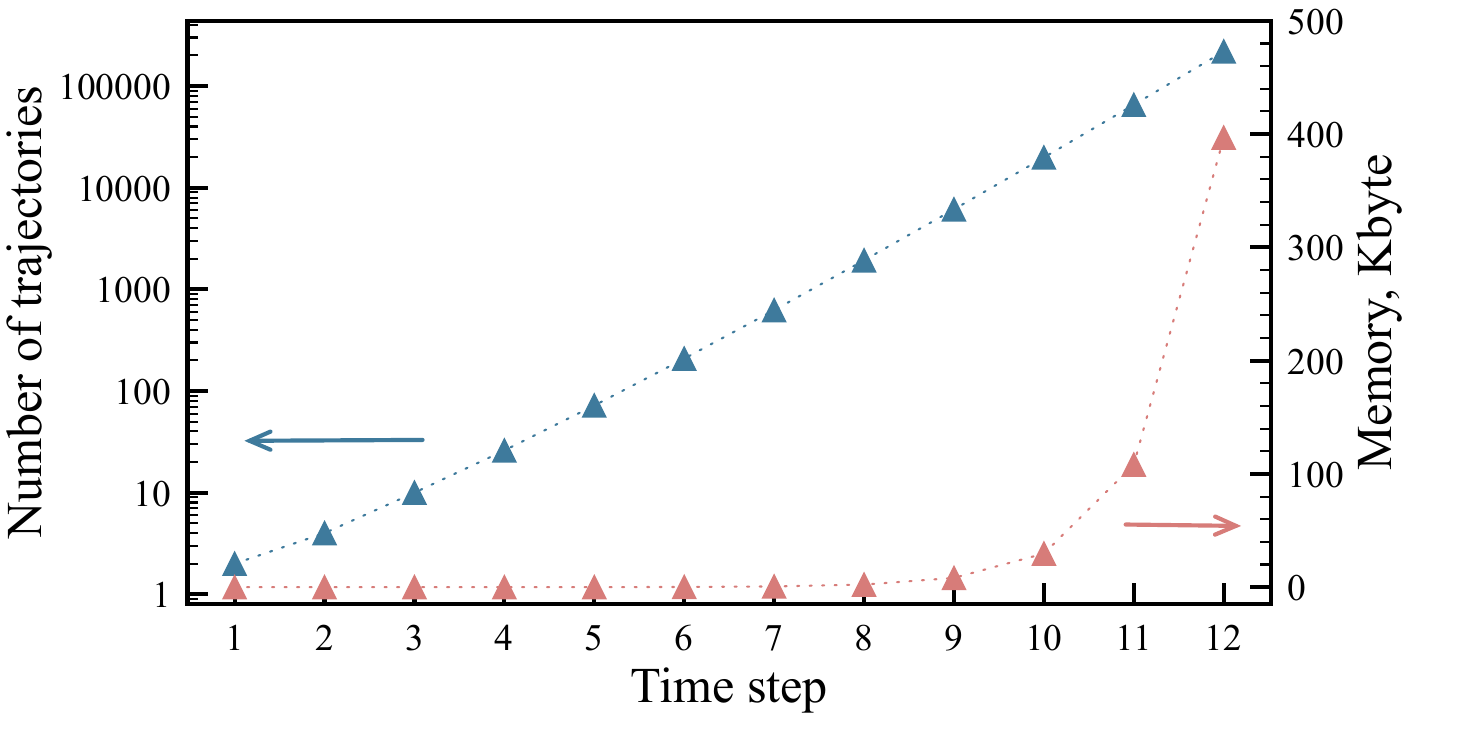}
  \caption{Number of trajectories (blue triangles) formed at different time steps of the evolution of the 50:50 loop-based BS. Lower bound of classical memory size (red triangles) needed for storing these trajectories.}
  \label{memory}
\end{figure}

The calculated Shannon entropies of the event series (Fig.~\ref{Shannon_entropy}) enable estimating classical memory resources needed to store the trajectories for theoretical characterization of the conditional wave functions formed during optical $\sf reset$ evolution, which is an integral part of dynamical quantum computing. Considering the case of the ideal 50:50 loop-based interferometer as the prominent example we define the total number of trajectories formed at each time step (Fig.~\ref{memory}) and on this basis we quantify the memory size one needs to reserve. Both the number of trajectories and memory size grow rapidly with increasing the time steps, for instance, at $t=12$ they are 218990 and 397 KBytes ($\approx 218990 \times 14.5 {\rm bits}$), respectively. Nowadays, an electronic device with memory of about 396 KBytes looks pretty ordinary. However, the main difficulty here is to implement a scheme for coding raw measurement data that allows such a high level of information compression in according with the Shannon's theory. If we assume that the loop beam splitter can generate only the events 1, 2, 3, or 4, and that 2 bits are used to encode each of them, then we need 24 bits to store a 12-step trajectory, which is larger than the Shannon estimate of 14.5 bits. In this respect the data lossless compression algorithms such as the Lempel-Ziv 77 coding algorithm (LZ77) \cite{LZ77} provide a bound for the Shannon entropy \cite{compress1,compress2}. Naturally, loop-based devices that feature multiple output modes with detectors will require an appropriate scaling of the classical resources.

\section{Incomplete information about measurement results}
Given that creating detectors with photon number resolution represents a huge technological problem and only a few realizations of such devices can be found in literature \cite{PND1,PND2, PND3}, it is important to discuss the approach we develop in this work from perspective of processing data generated by simpler time-bin optical devices without distinguishing the output states with respect to photon number. In this case, a history of measurement results in the mode $c$ at the given the $t$-th time step  can be represented as a bitstring, $y_{t} = (x_1,...,x_{t}) $ where $x_{\tau} = 0$ and $x_{\tau} = 1$ correspond to no-click and click events, respectively. 

Since there is no information on the exact number of output photons at each time step, one cannot define the occupation of the loop part and, as a result, decrease uncertainty in measurement results at the next step. This is supported by the calculated values of the event-history mutual information, Eq.~\eqref{MI_classical} for the 50:50 beam splitter that are strongly suppressed with respect to the results presented in Fig.~\ref{Class_mutual_info}.  For instance, $J(y_{11}:x_{12})$ is equal to 0.695 with distinguishing the number of photons and 0.017 without. A similar suppression is observed for calculated event-history correlation functions.

\end{document}